\documentclass[aps,prl,twocolumn,superscriptaddress]{revtex4}

\usepackage{psfig}
\begin{document}
\psfigurepath{.}

\title{Anomalous critical behavior near the
quantum critical point of a hole-doped La$_2$CuO$_4$}

\author{Y. Chen}
\author{Wei Bao}
\affiliation{Los Alamos National Laboratory, Los Alamos, NM 87545}
\author{J.E. Lorenzo}
\affiliation{CNRS, BP 166X, F-38043, Grenoble, France}
\author{A. Stunault}
\affiliation{Institut Laue-Langevin, BP 156, F-38042, Grenoble, France}
\author{J.L. Sarrao}
\affiliation{Los Alamos National Laboratory, Los Alamos, NM 87545}
\author{S. Park}
\affiliation{NIST Center for Neutron Research, National Institute of Standards and Technology, Gaithersburg, MD 20899}
\affiliation{Dept.\ of Materials Science and Engineering, University of
Maryland, College Park, MD 20742}
\affiliation{HANARO Center, Korea Atomic Energy Research Institute,
Daejeon 305-600, Korea}
\author{Y. Qiu}
\affiliation{NIST Center for Neutron Research, National Institute of Standards and Technology, Gaithersburg, MD 20899}
\affiliation{Dept.\ of Materials Science and Engineering, University of
Maryland, College Park, MD 20742}
\date{\today}

\begin{abstract}
The Landau-Ginzburg-Wilson paradigm for critical phenomena is spectacularly
successful whenever the critical temperature is finite and all fluctuation
modes, with characteristic energies much smaller than $k_B T_C$, obey
classical statistics. In zero-temperature quantum critical phenomena,
classical thermal fluctuations are replaced by zero-point quantum fluctuations
and quantum-mechanical generalization of the Landau-Ginzburg-Wilson paradigm
has been a central topic in condensed-matter physics.
In this neutron-scattering study on spin fluctuations near 
the quantum critical point induced by hole-doping in La$_2$Cu$_{1-x}$Li$_x$O$_4$
($0.04 \le x \le 0.1$), the phase boundary for quantum crossover expected from 
the generalized quantum theory for critical phenomena was 
observed for the first time. Furthermore, critical exponent and 
scaling function become anomalous near the quantum critical point, 
which has not been expected in current theories.
\end{abstract}

\pacs{}

\maketitle

\begin{figure}
\centerline{
\psfig{file=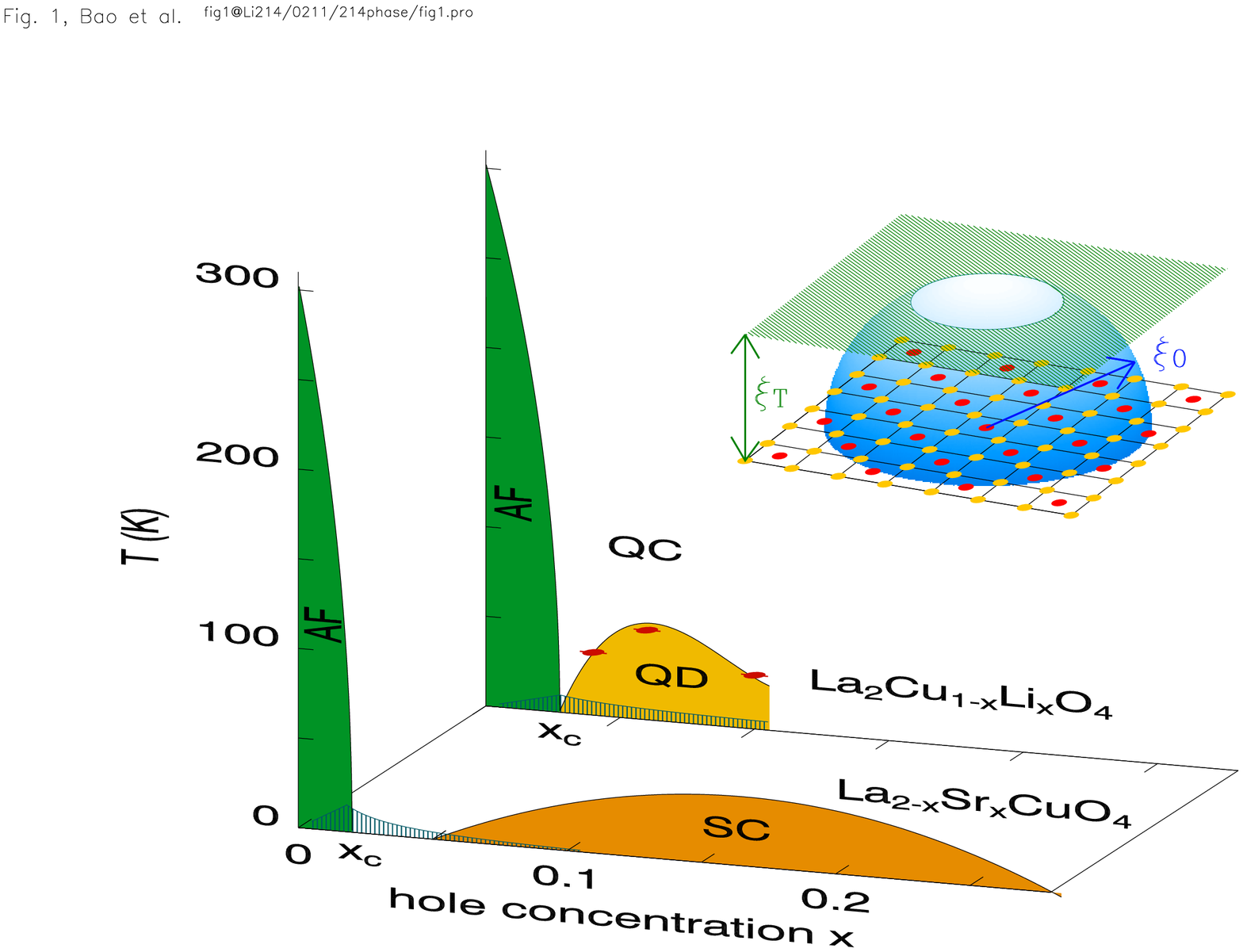,width=1.\columnwidth,angle=90,clip=}}
\vskip -7ex 
\caption{\label{fig1} (color)
Phase diagrams for Sr and Li-doped La$_2$CuO$_4$ where increasing
hole concentration suppresses the antiferromagnetic (AF) phase
at the quantum critical point $x_c$
and drives the Sr doped material superconducting (SC). The line filled
area denotes spin-glass phase. We observe the crossover at $T_X$ 
(red symbols) between the $E/T$ scaling (QC) and the $E/\Gamma_0$ scaling 
(QD) regimes in Li-doped La$_2$CuO$_4$. 
The inset illustrates the
(2+1)-dimensional space formed by the
CuO$_2$ plane (Cu: red circles, O: yellow circles) and the 
quantum $\xi_T$. The blue hemisphere indicates the extent of the
magnetic correlation length at zero temperature, $\xi_0$.
}
\end{figure}

Interesting phenomena, including high-$T_C$ superconductivity, occur
in laminar cuprates below a temperature $T\ll J/k_B\approx 1500$~K,
where $J$ is the magnetic exchange energy
between the nearest-neighbor spins in the CuO$_2$ plane.
In this temperature range, quantum fluctuations become dominant 
and add one extra dimension, $\xi_T \equiv \hbar c/k_BT$, 
where $c$ is the spin-wave velocity, to the 2-dimensional (2D) CuO$_2$ square 
lattice\cite{ja_hertz,2dheis} (Fig.~\ref{fig1}). 
At the zero temperature, a quantum phase transition from the  
antiferromagnetic (AF) phase to a paramagnetic one occurs at a critical
hole doping, a.k.a.\ quantum critical point (QCP), 
$x_c$, in the cuprates. Thus, for doping $x>x_c$ on the paramagnetic side of the QCP,
magnetic correlation length at $T=0$, $\xi_0$, is finite.
Generalizing the Landau-Ginzburg-Wilson paradigm for critical phenomena in
classical systems to the quantum spin system
in the (2+1)-dimensional space, the shorter of $\xi_T$ and $\xi_0$ 
sets the long-wavelength 
cutoff for spin fluctuations and is expected to determine the universal
magnetic phenomena with universal critical exponents\cite{2dheiqc}.
The $E/T$ scaling in spin dynamics, observed in diverse
samples of hole-doped La$_2$CuO$_4$
using Ba, Sr or Li dopant, with superconducting or insulating
ground state, and with commensurate or incommensurate
dynamic spin correlations\cite{la2smha,la2keimer,la2gas,bao02c},
can be understood as the physical consequence of 
$\xi_T <\xi_0$ at $T> T_X\equiv \hbar c/k_B\xi_0$.

In this neutron scattering study on quantum spin dynamics in
hole-doped La$_2$CuO$_4$, we determine the doping dependence of the 
crossover temperature $T_X$, below which the $E/T$ scaling breaks down and
crosses over to a constant-energy scaling, consistent with $\xi_0 <\xi_T$. 
The most interesting result of this study, however, is that 
the critical exponent $a$ in the $E/T$ scaling\cite{2dheiqc}
changes from an expected $a\approx 1$ to an anomalous $a\approx 0.65$ 
when doping is reduced toward the AF QCP at $x_c$. 
The change of the critical exponent is inconsistent with the classical 
universality of critical exponents. In analog to a recent example in 
heavy fermion metal\cite{qcp_as,qcp_sqm}, the substantial reduction of
the $a$ exponent in the 2D spin system suggests extra physics in quantum
critical phenomena which invalidates the 
classical Landau-Ginzburg-Wilson paradigm\cite{qcp_ts}.

We choose Li-doped La$_2$CuO$_4$ single crystals for this study (Table I). 
\begin{table}[bht]
\caption{Doping $x$ of La$_2$Cu$_{1-x}$Li$_x$O$_4$
and main experimental quantities determined in this study}
\begin{ruledtabular}
\begin{tabular}{c|ccc}
$x$ & 0.04 & 0.06 & 0.1\\
\hline
$T_X$ (K) & 35 & 50 & 30\\
$a$ & 0.65$\pm$0.06 & $\approx$1 & $\approx$1 \\
$b$  & 0.21$\pm$0.01 & 0.18$\pm$0.01 & 0.28$\pm$0.01\\
\end{tabular}
\end{ruledtabular}
\end{table}
Isotopically enriched $^7$Li (98.4\%) was used to reduce neutron 
absorption of natural Li. The crystals have orthorhombic $Cmca$ symmetry 
in the temperature range of this study and lattice parameters were reported 
in \cite{Li214}. Analogous to Sr-doped La$_2$CuO$_4$\cite{fchou,nagano,muchn}, 
the AF phase is suppressed rapidly
by 3 percent of holes and a spin-glass transition occurs
below 10~K on both sides of the QCP\cite{Li214phs,Li214phs2} (Fig.~\ref{fig1}). 
In addition, similar spin dynamics following the $E/T$ scaling have been
observed\cite{la2smha,la2keimer,la2gas,bao02c}. 
To investigate the intrinsic spin dynamics of the 2D quantum antiferromagnet, 
the suppression of the $d$-wave superconductivity
by Li dopants as potential scatterers in the CuO$_2$ plane\cite{thy_pl}
is crucial. This
allows measurement of the momentum and energy-dependent 
dynamic magnetic structure factor, $S({\bf q},E)$, at low temperature 
without interference by the superconducting 
gap as seen 
in Sr or Ba-doped La$_2$CuO$_4$\cite{la2sym,la2syma}.
To achieve required energy resolution,
the cold neutron triple-axis spectrometer IN14 at the Institut
Laue-Langevin with $E_f$=5~meV (for $x$=0.1),
and SPINS at the National Institute of Standards and Technology Center 
for Neutron Research with $E_f$=3.7~meV (for $x$=0.04) were used.
The scattered neutron beam was filtered using cooled Be or BeO to remove
higher order neutrons.
The dynamic spin correlations which can be resolved in this cold
neutron inelastic scattering study do not participate 
in the spin-freezing process\cite{bao02c}. We will report 
the  minority spin-glass phase elsewhere.

\begin{figure}
\centerline{\psfig{file=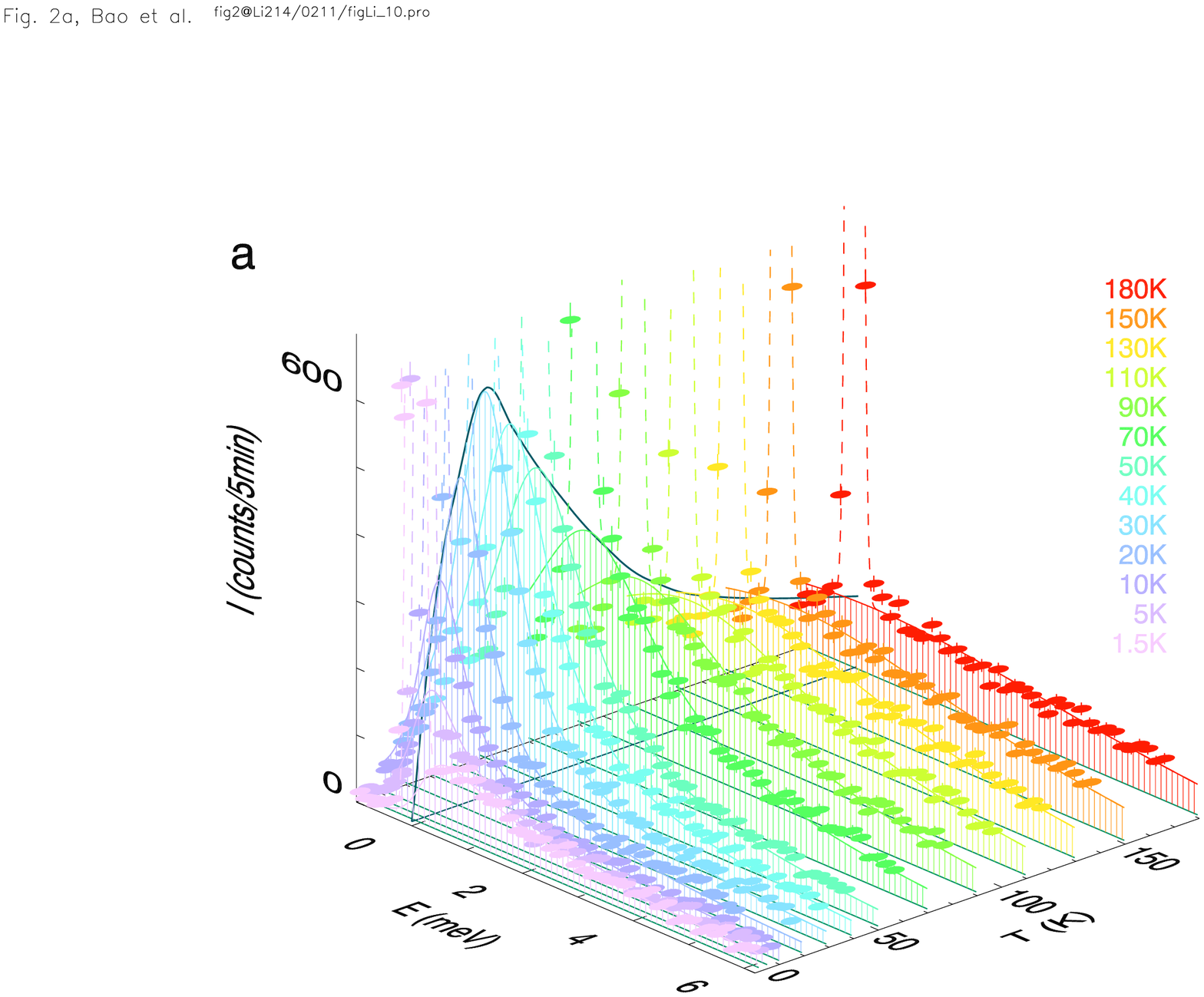,width=1.\columnwidth,angle=90,clip=}}
\vskip -10ex
\centerline{\psfig{file=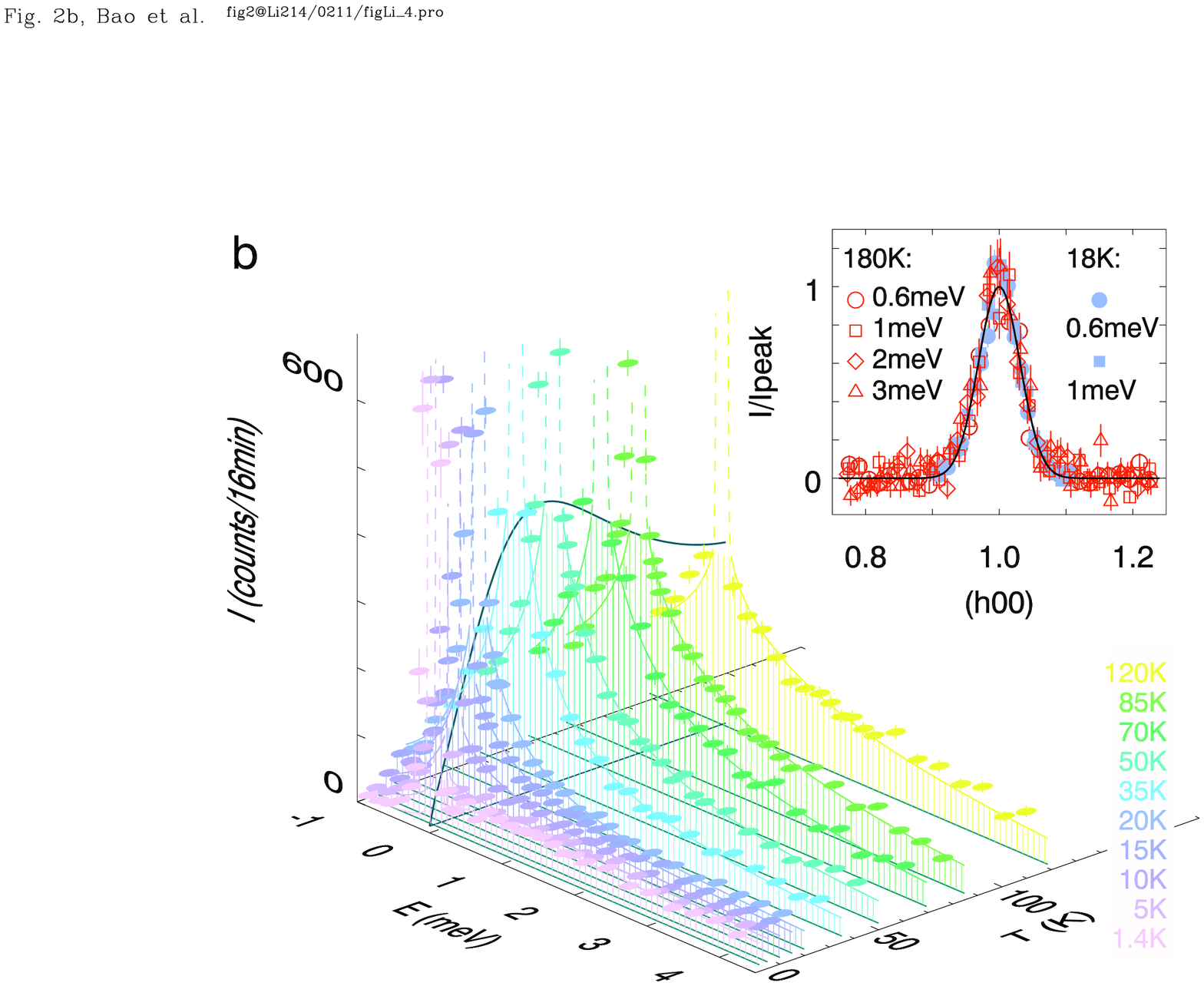,width=1.\columnwidth,angle=90,clip=}}
\caption{\label{fig2}
(color) Measured  dynamic magnetic structure factor $S({\bf Q},E)$ 
as a function of $E$ and $T$
at {\bf Q}=(100)
in the $Cmca$ orthorhombic notation which corresponds to the ($\pi,\pi$)
point of the CuO$_2$ square lattice. {\bf a}, Li doping $x$=0.1. 
{\bf b}, $x$=0.04. The blue curve at $E$=0 traces $S({\bf Q},E)$
at the low energy limit as a function of temperature.
Inset: normalized $S({\bf q},E)$ as
a function of {\bf q} at indicated $E$ and $T$ for $x$=0.1.
}
\end{figure}

Dynamic spin correlations in Li-doped La$_2$CuO$_4$ are finite-sized 
fluctuating versions of the chessboard-like antiferromagnetic order in pure 
La$_2$CuO$_4$\cite{bao99a}. Intensity of $S({\bf q},E)$ 
concentrates in reciprocal space in rods intercepting the planar 
Brillouin zone of the CuO$_2$ square lattice at commensurate Bragg points 
of the ($\pi,\pi$)-type.
Scans at various $E$ and $T$ across the rod 
at {\bf Q}$\equiv$($\pi,\pi$) are shown in the inset to Fig.~\ref{fig2} 
for Li doping $x$=0.1. Note that {\bf Q}=(100) in the orthorhombic notation. 
The scans have been normalized by the peak intensities so that
the invariance of the in-plane peak width in this work using coarse
{\bf q} resolution can be easily seen.
Reflecting the two-dimensional nature of spin correlations, 
$S({\bf q},E)$
is flat in the out-of-plane direction\cite{bao02c}.
Thus, $\int d{\bf q} S({\bf q},E)$ is simply proportional to
the value of $S({\bf Q},E)$ in the $T$
and $E$ range of this study, and the imaginary part of the local magnetic 
response function
$
\chi''(E)\equiv \left(1-e^{-E/k_BT}\right) 
\int d{\bf q} S({\bf q},E)
$
is proportional to
$ \left(1-e^{-E/k_BT}\right) S({\bf Q},E) $.
Measured $S({\bf Q},E)$  
as a function of $E$ and $T$ is shown in 
Fig.~\ref{fig2}a and b for $x$=0.1 and 0.04, respectively.
Throughout this paper, a unique color is assigned to each
temperature for data points in the figures.
The peak at $E$=0 (dashed curves) contains incoherent, elastic and quasielastic
scattering processes within the energy resolution, $\Delta_{\rm res}$, 
of the spectrometer and is not the subject of this work. 
For the resolvable dynamic signal, $S({\bf q},E)$ with 
$|E|> \Delta_{\rm res}$, background (straight green lines)  
is determined by measurement at {\bf q} far away from 
the peak of $S({\bf q},E)$ (refer to the inset to Fig.~\ref{fig2}), or at $E\ll -k_BT$
where the Bose factor for spin fluctuation modes approaches zero.

\begin{figure}
\centerline{\psfig{file=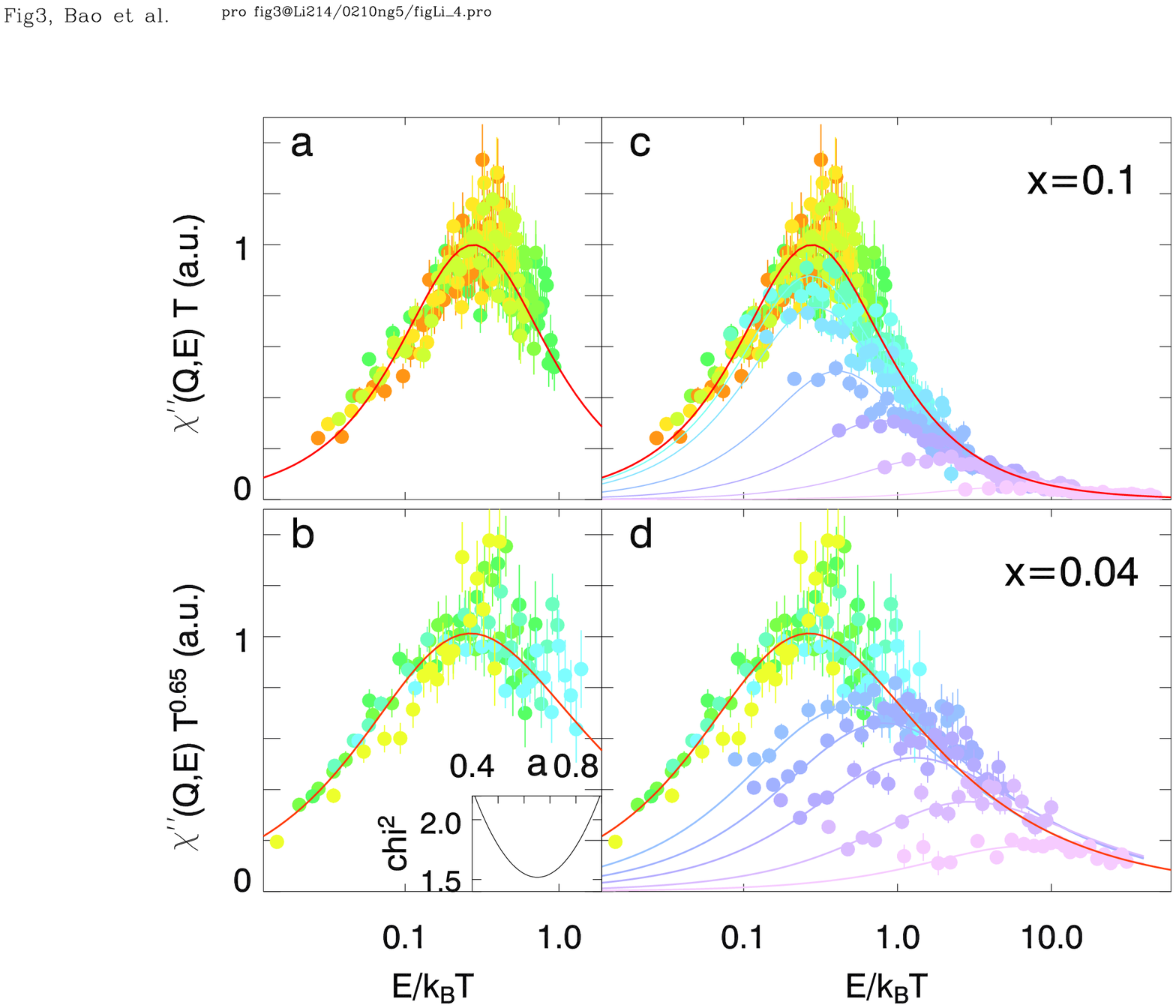,width=1.\columnwidth,angle=90,clip=}}
\vskip -3ex
\caption{\label{fig3}
(color) Scaling plots of the magnetic response function.
{\bf a}, {\bf b}, The $E/T$ scaling at high temperature for
$x$=0.1 and 0.04, respectively. 
Inset in {\bf b} shows the goodness-of-fit for the critical 
exponent $a$=0.65 for $x$=0.04.
{\bf c}, {\bf d}, Departure
from the $E/T$ scaling at low temperature. 
The solid line in {\bf a} and {\bf c} is Eq.~(\ref{eq0}) using
Eq.~(\ref{eq1}) as the scaling function, in {\bf b} and {\bf d}
using Eq.~(\ref{eq2}) as the scaling function. Color of symbols
denotes temperature as in Fig.~\ref{fig2}.
}
\end{figure}

To demonstrate the $E/T$ scaling in magnetic response 
function:
\begin{equation}
\chi''(E)T^{a}=A f(E/bk_BT), \label{eq0}
\end{equation}
where $a$ is a critical exponent\cite{2dheiqc},
$A$ a constant, $f(y)$ a scaling function defined to peak at $y$=1, 
and $b=E_p/k_BT$ 
relates the peak energy of $\chi''(E)$ to thermal energy, 
experimental data of $\chi''(E)$ above $T_X$, 
renormalized by $T^{-a}$, are plotted
semilogarithmically as a function of $E/k_BT$ 
for $x$=0.1 and 0.04 in Fig.~\ref{fig3}a and b, respectively.
For $x$=0.1, identical to the case for 
$x$=0.06\cite{bao02c}, $a\approx 1$ and the scaling function is
\begin{equation}
f_1(y)=y/(1+y^2). \label{eq1}
\end{equation}
The peak position $E_p/k_BT$=$b=0.28\pm 0.01$ in Fig.~\ref{fig3}a 
is larger than $b$=0.18 for $x$=0.06.
For Ba or Sr-doped La$_2$CuO$_4$, the exponent $a$ has been 
determined only for $x$=0.14 and it is also $a=0.94\pm 0.06\approx 1$\cite{la2gas}.
Thus for $x$$\ge$0.06, hole-doped La$_2$CuO$_4$ has a normal 
critical exponent $a$$\approx$1, consistent with a Curie's law for
the staggered magnetic susceptibility.

For Li doping $x$=0.04 which is close to the AF QCP of 
$x_c$$\approx$$0.03$\cite{Li214}, 
a fractional $a$=$0.65\pm 0.06$ is required for the data
to collapse onto a single scaling curve (Fig.~\ref{fig3}b).
Previously, in heavy fermion alloy CeCu$_{5.9}$Au$_{0.1}$, which is also
near an AF QCP, single-crystal 
neutron scattering data are collapsed onto the $E/T$ scaling
with $a$=$0.74\pm 0.1$\cite{qcp_as}.
The same procedure as in Ref.~\cite{qcp_as}, 
which is unbiased with respect to the
selection of $f(y)$, is used to optimize $a$ for $x$=0.04,
and the goodness-of-fit represented by the chi-square statistical
test is shown in the
inset to Fig.~\ref{fig3}b.
The spectrum in Fig.~\ref{fig3}b is broader than that in
Fig.~\ref{fig3}a and cannot be described by Eq.~(\ref{eq1}).
We adopt empirically the scaling function for $x$=0.04, 
\begin{equation}
f_2(y)=sgn(y)\,\, y_1^u/(1+y_1^v), \label{eq2}
\end{equation}
where $y_1$$\equiv$$ |y| [(v-u)/u]^{1/v}$ with $u$=0.77$\pm$0.02 and
$v$=1.33$\pm$0.02 as the best fits.
The parameter $b=0.21\pm 0.01$. Note that for $|y| \ll 1$, 
$f_2(y)\sim sgn(y)|y|^u$ is also anomalous and does not conform to
the usual analytic form $\chi''(E)\sim E$ at small $E$\cite{2dheiqc}.

The heavy fermion metal CeCu$_{5.9}$Au$_{0.1}$ ($z$=2 and dimension $d$$>$$2$)
and our $x$=0.04 cuprate sample ($z$=1 and $d$=2) belong to different 
universality classes with the effective dimension ($d+z$) above and below the
upper critical dimension, respectively\cite{ja_hertz,qpt_ss}.
However, they are both close to an AF QCP and share
within error bars an anomalous $a$ exponent substantially smaller than unity. In our study of doping dependence,
furthermore, hole-doped La$_2$CuO$_4$ recovers the normal $a\approx 1$ and
analytic $\chi''(E)$ for $x\ge 0.06$. These changes in critical
behavior are very unusual.
Anomalous $a$ exponent for the heavy fermion 
metal CeCu$_{5.9}$Au$_{0.1}$ is recently explained by including
local degrees of freedom in the Kondo screening, in addition to traditional
long-wavelength magnetic fluctuations, in the
antiferromagnetic quantum phase transition\cite{qcp_sqm}.
Closer to our case of doped cuprates near the AF QCP,
currently, ``deconfined'' degrees of freedom associated with fractionalization of 
magnetic order parameter in pure  
2D antiferromagnet are being explored theoretically\cite{qcp_ts}. 
Like Ref.~\cite{qcp_sqm}, this theory goes beyond the Landau-Ginzburg-Wilson 
paradigm, which focuses on long-wavelength fluctuations 
of the order parameter in critical phenomena.
Whether including these additional ``deconfined'' degrees of freedom in 
the critical theory for the QCP would, similar to the case for heavy fermion 
metals\cite{qcp_sqm}, produce the anomalous $a$ exponent and $f_2(y)$ 
reported here for cuprates is unknown. 
Nevertheless, our observation of anomalous critical phenomena near the QCP
in another class of correlated electronic material suggests the generality
of the phenomena, and would certainly encourage the budding
research interest on subtle quantum interference effects near a QCP.

\begin{figure}
\centerline{\psfig{file=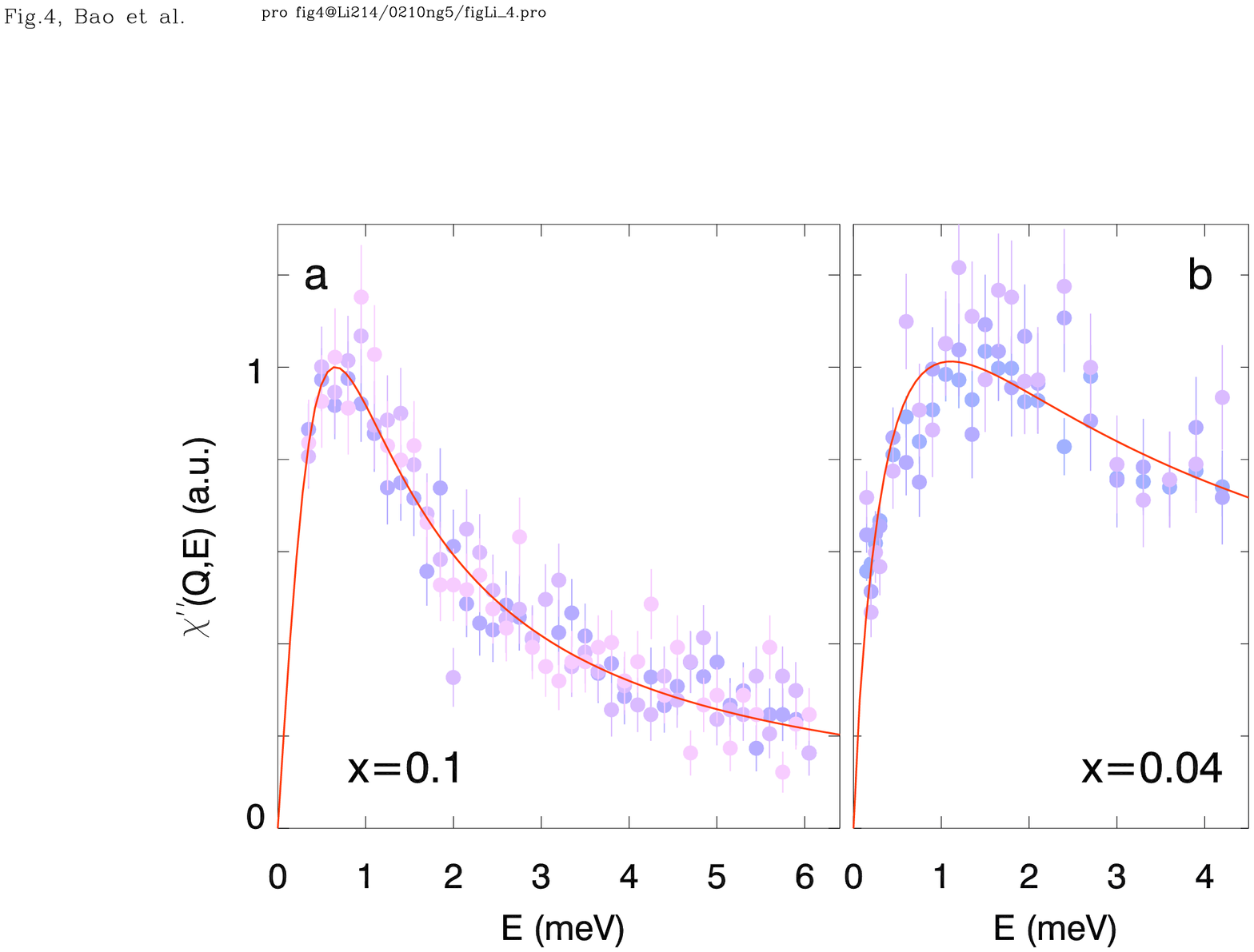,width=1.\columnwidth,angle=90,clip=}}
\vskip -8ex
\caption{\label{fig4}
(color) Temperature invariance of the magnetic response function
at $T<T_X$. 
{\bf a}, The doping $x$=0.1. The solid line is Eq.~(\ref{eq3}) using
Eq.~(\ref{eq1}) as the scaling function.
{\bf b}, $x$=0.04. The solid line is Eq.~(\ref{eq3}) using
Eq.~(\ref{eq2}) as the scaling function.
Color of symbols
denotes temperature as in Fig.~\ref{fig2}.
}
\end{figure}

In Fig.~\ref{fig3}c and d, low temperature data are added to the
scaling plots for $x$=0.1 and 0.04, respectively. 
Since the peak position shifts to the right with decreasing temperature,
the $E/T$ scaling cannot be satisfied. Instead, as shown in
Fig.~\ref{fig4}a and b, below $T_X$, the local magnetic response function 
$\chi''(E)$ is nearly temperature independent. 
The curves in the figures are
\begin{equation}
\chi''(E)=B f(E/\Gamma_0), \label{eq3}
\end{equation}
where $B$ is a constant, and the same pair of scaling functions,
$f_1(y)$ and $f_2(y)$, are used in Eq.~(\ref{eq3}) with 
$\Gamma_0$=0.66$\pm$0.01 and 1.11$\pm$0.03~meV for $x$=0.1 and 0.04, respectively. 
Note that there is no detectable gap in $\chi''(E)$ in
Fig.~\ref{fig4} for $x$=0.1 and 0.04, as for $x$=0.06
reported previously\cite{bao02c}. In some theoretical models of nonrandom quantum antiferromagnets, a quantum spin liquid has a gap, $\Delta\sim \hbar c/\xi_0$, 
near $E$=0 in $\chi''(E)$ below $T_X$\cite{qpt_ss}.
The gapless $\chi''(E)$ observed in our samples is obtained in these theories
with strong damping of the spin liquid by doped holes,
$\Gamma \approx \Delta$\cite{2dheiz2,2dheisu}. Partial filling
of the Haldane gap in 1D quantum spin liquid by doped holes has been 
reported lately\cite{ybno_gx}.
In other theoretical models of quantum antiferromagnets, several types of
gapless spin liquids exist without impurity 
scattering\cite{q_wen}. Classification of our observed 2D gapless spin liquid
in hole-doped cuprates would be an interesting theoretical task.

The crossover temperature, $T_X$, from the $E/T$ scaling to the 
$E/\Gamma_0$ scaling regime
can be conveniently located by the peak in temperature
dependence of $S({\bf Q},E)$ 
at the $E$$\rightarrow$$0$ limit\cite{bao02c}, since at high temperature,
$
S({\bf Q},E$$\rightarrow$$0)\propto T^{-a} 
$ 
from Eq.~(\ref{eq0})-(\ref{eq2}), and at low temperature
$ 
S({\bf Q},E$$\rightarrow$$0)\propto T 
$ 
from Eq.~(\ref{eq1})-(\ref{eq3}).
The blue curves at $E$=0 in Fig.~\ref{fig2}a and b are 
$S({\bf Q},E$$\rightarrow$$0)$ for $x$=0.1 and 0.04, respectively.
The extracted $T_X$ from Fig.~\ref{fig2}, together
with previously determined value for $x$=0.06\cite{bao02c}, are shown in 
the phase diagram in Fig.~\ref{fig1} and are listed in Table~I. 
Under the dome-shaped phase boundary,
spin fluctuation energy, which would approach zero according to 
the $E/T$ scaling, saturates at a finite $\Gamma_0$. Thus the crossover
at $T_X$ keeps spin fluctuations energetic at low temperature,
which might be essential for the high value of $T_C$ if superconductivity
in cuprates is mediated magnetically by spin fluctuations.

Experimental search for the $E/T$ scaling was initially stimulated by
the marginal Fermi liquid phenomenology\cite{bibmfl,la2smha,la2keimer}.
The subsequent observation of the scaling in insulating 
sample\cite{bao02c} supports
the QCP as the common mechanism. 
The breakdown of the $E/T$ scaling below a
dome-shaped crossover boundary reported here excludes the existence of additional 
magnetic QCPs in the investigated doping range. The gapped
quantum spin liquid predicted below $T_X$ for 
nonrandom 2D antiferromagnets does not survive 
in our doped cuprates. This observation disfavors those theories which
rely on the robustness of the gapped quantum spin liquid against doping
for high-$T_C$ superconductivity in cuprates\cite{subir_rev}.
While the application of the classical Landau-Ginzburg-Wilson paradigm 
for critical phenomena to quantum critical phenomena in cuprates
naturally explains the universal occurrence of the $E/T$ 
scaling in hole-doped La$_2$CuO$_4$ and the crossover to 
the finite magnetic-energy
regime at low temperature\cite{2dheis,2dheiqc}, the anomalous 
critical exponent and scaling function near the QCP reported here have yet
to be understood, probably by going beyond the Landau-Ginzburg-Wilson 
paradigm and taking into account additional quantum interference
effects existing near the zero temperature\cite{qcp_sqm,qcp_ts}.

We thank Q.M. Si, A. Zheludev, S.-H. Lee, C. Broholm, S. Sachdev, 
A.V. Chubukov, F.C. Zhang, C.M. Varma, X.G. Wen, A. Balatsky, A. Abanov, L. Yu, 
Z.Y. Weng and Y. Bang for useful discussions. 
SPINS at NIST is supported partially by NSF. Work at LANL is
supported by U.S. Department of Energy.

\end{document}